\documentclass[conference]{IEEEtran}
\usepackage[utf8]{inputenc}

\usepackage{booktabs}

\usepackage{comment}

\usepackage{svg}

\usepackage{bm}

\usepackage{url}

\usepackage{cite}

\usepackage{xcolor}

\usepackage{stackrel}

\usepackage{todonotes} % great for draft annotations
\ifCLASSOPTIONcompsoc
  \usepackage[caption=false,font=normalsize,labelfont=sf,textfont=sf]{subfig}
\else
  \usepackage[caption=false,font=footnotesize]{subfig}
\fi

\usepackage{lipsum}
\usepackage{algorithm}
\usepackage{siunitx}

\usepackage{mathtools}
\usepackage{csquotes}
\usepackage{xcolor}  % For nice colorfull texts (e.g. marking unfinished texts)

\usepackage{pgf} % for pgf plots from matplotlib.

\usepackage{bbm}
\usepackage{amsbsy}
\usepackage{amssymb}
\usepackage{amsmath}
\usepackage{exscale} %% Fuer 11pt und 12pt-fonts korrekte integral-und
\usepackage{icomma} %%Benutze Komme als Dezimaltrennzeichen, nurLeerschritt, wenn mit " " eingegeben
\usepackage{mathrsfs} %% Mahtematische Schreibschrift mit \mathscr{}
\usepackage{array} %% für Tabellen
\usepackage{multirow} %% für Tabellen
\usepackage{listings} %% für Sourcecode
\usepackage{relsize}
\usepackage[inline, shortlabels]{enumitem}
\usepackage{dsfont}
\usetikzlibrary{positioning}
\usetikzlibrary{external}
\tikzsetexternalprefix{tikz_fig/}

\newcommand{\argmax}{\text{arg}\,\text{max}}

\begin{document}

\title{Deep Reinforcement Learning for Wireless Resource Allocation Using Buffer State Information}

\author{\IEEEauthorblockN{Eike-Manuel Bansbach, Victor Eliachevitch, and Laurent Schmalen}
\IEEEauthorblockA{Communications Engineering Lab, Karlsruhe Institute of Technology (KIT), 76187 Karlsruhe, Germany\\ (e-mail: \texttt{e.bansbach@kit.edu})}
}


\maketitle

\begin{abstract}
As the number of user equipments (UEs) with various data rate and latency requirements increases in wireless networks, the resource allocation problem for orthogonal frequency-division multiple access (OFDMA) becomes challenging.
In particular, varying requirements lead to a non-convex optimization problem when maximizing the systems data rate while preserving fairness between UEs.
In this paper, we solve the non-convex optimization problem using deep reinforcement learning (DRL).
We outline, train and evaluate a DRL agent, which performs the task of media access control scheduling for a downlink OFDMA scenario.
To kickstart training of our agent, we introduce mimicking learning. 
For improvement of scheduling performance, full buffer state information at the base station (e.g. packet age, packet size) is taken into account.
Techniques like input feature compression, packet shuffling and age capping further improve the performance of the agent. 
We train and evaluate our agents using Nokia's wireless suite and evaluate against different benchmark agents.
We show that our agents clearly outperform the benchmark agents.

\end{abstract}


% \fi
\IEEEpeerreviewmaketitle

\section{Introduction}\label{chap:intro}
As Internet of things applications, such as connected cars, drones and augmented reality,  boost the massive growth of data traffic, the challenges to fulfill latency and reliability requirements are unprecedented \cite{bennisUltrareliableLowLatencyWireless2018}. 
Dependent on the service, e.g. conversational voice, conversational video or web browsing, the guaranteed bit rate (GBR) as well as the packet delay budget (PDB) differ \cite{valcarceTimeFreqResourceAllocationv0Environment2020}.
Since different quality of service (QoS) classes aim for different GBRs and PDBs, increasingly different QoS requirements of the user equipments (UEs) need to be taken into account by an acceptable network design \cite{shamsSurveyResourceAllocation2014}.

Orthogonal frequency-division multiple access (OFDMA) is a widely used media access scheme in wireless communications, as it has a high resistance to frequency-selective fading, while at the same time enabling a high flexibility for radio resource allocation \cite{shamsSurveyResourceAllocation2014}. 
To set up an OFDMA system that fulfills the differing requirements of UEs denoted by their QoS class, a well devised resource allocation algorithm is necessary. 
For the downlink scenario, an efficient centralized scheduling of the available OFDMA subcarriers at the base station (BS) can ensure that the minimum required QoS is provided \cite{shamsSurveyResourceAllocation2014}.

For UEs of a single QoS class with time-invariant data rate requirements, the OFDMA resource allocation problem can be formulated as a convex optimization problem with the task to maximize the data rate of the system.
Depending on the constraints of the optimization problem, there is a tradeoff between maximizing the total throughput and traffic fairness of different UEs \cite{shamsSurveyResourceAllocation2014}.
Water filling is throughput maximizing but unfair, since resources are allocated to UEs with good channels and UEs with bad channels may not get allocated any resource \cite{chengGaussianMultiaccessChannels1993}.
At the cost of data rate, max-min fairness provides fair shared resources among all UEs \cite{rheeIncreaseCapacityMultiuser2000}.
A tradeoff of both is weighted proportional fairness (PF) \cite{kellyRateControlCommunication1998}. 
However, these approaches are not capable of dealing with variable QoS requirements among UEs and time-variant data rate requirements, which results in a non-convex optimization problem \cite{shamsSurveyResourceAllocation2014}.
To solve the allocation task for UEs of different QoS classes, \cite{Utility_max-min} introduces a utility function which models the application layer QoS.
This results in a convex optimization problem again \cite{huangJointSchedulingResource2009}. 

For more flexibility, \cite{wangDeepReinforcementLearning2019} proposes to model the allocation problem as a Markov decision process (MDP) and solve it by deep reinforcement learning (DRL), since DRL facilitates optimization over high-dimensional data.
In comparison to reinforcement learning (RL), DRL takes advantage of deep neural networks (DNNs) and thus improves learning speed and performance in the training process \cite{luongApplicationsDeepReinforcement2019}.
In \cite{wangDeepReinforcementLearning2019}, a DRL agent for a downlink OFDMA resource allocation scenario is trained, where for each time step all the resources are allocated to a single UE. 
The agent knows the instantaneous data rate and average data rate for each UE, while all UEs have the same QoS and the BS buffer is always filled with packets to transmit.
An \emph{expert learning} method is introduced that boosts the training of the DRL agent. 
In \cite{al-tamLearnScheduleLEASCH2020} and \cite{Xu20}, a DRL agent which allocates the frequency resources of the OFDMA system to different UEs is proposed.
Time-varying data rates of UEs are simulated by unoccupied BS buffer slots.
While the approach of \cite{al-tamLearnScheduleLEASCH2020} includes buffer state information, solely indicating whether a packet for a certain UE is in line or not, \cite{Xu20} extends buffer state information to the waiting time of the next packet to transmit for every UE as well as the spare space in the buffer.  
Both outperform PF scheduling, \cite{al-tamLearnScheduleLEASCH2020} for 4 and 8 UEs with alike QoS and \cite{Xu20} for 5 UEs with alike QoS.
In \cite{Lee20}, a DRL agent is designed which is able to adapt to a variable number to UEs after training, ranging from 4 to 20 UEs. 
However, the adaptive agent isn't able to outperform an agent specifically trained for a fixed number of UEs.

While previous work was limited to a small number of UEs with alike QoS and limited BS buffer state information, we propose a method to solve the OFDMA resource allocation task for an increased number of UEs with varying QoS and full BS buffer state information, e.g. the size and the age of packets.
Adapting expert learning \cite{wangDeepReinforcementLearning2019}, we kickstart our DRL agent, but then reduce the influence of the expert to enable our DRL agent to outperform the expert agent.  
By providing the age and size of every packet to the agent, more detailed buffer state information improves the trained DRL agent.
However, the input dimensionality to the DRL agent increases tremendously.
Inspired by autoencoders, we introduce novel encoder neural networks (ENNs), which apply feature extraction to compress buffer state information and therefore reduce the input dimension to the agent. 
To allow for generalization, we introduce an age capping technique.

We benchmark our agents against the open source ``Wireless Suite'' problem collection by Nokia \cite{NokiaWirelesssuite2021} with its \textit{TimeFreqResourceAllocation-v0} (TFRA) environment.
We show that our trained agents outperform the agents supplied by the ``Wireless Suite''.

\section{Resource Allocation Problem}\label{chap:RAP}
\subsection{Formulation of the Optimization Problem}
OFDMA resource allocation algorithms are typically classified by the objective of the underlying optimization problem.
While margin-adaptive schemes aim at minimizing the power consumption while complying with a set of fixed user requirements, rate-adaptive schemes aim at maximizing the total sum data rate over all UEs while satisfying power consumption constraints and other QoS requirements.
A common formulation of the rate-adaptive OFDMA optimization problem is given by \cite{shamsSurveyResourceAllocation2014}.
Having the set of $N$ subcarriers $\mathcal{N}=\{1,\ldots N\}$ and the set of $K$ UEs $\mathcal{K} = \{1,\ldots K\}$, the subset of subcarriers $\mathcal{N}_k \subseteq \mathcal{N}$ gets assigned to UE $k \in \mathcal{K}$.
Since each subcarrier can only be assigned to one UE the subsets of subcarriers $\mathcal{N}_k$ are pairwise disjoint.
$r_{kn}$ denotes the data rate that UE $k$ can achieve over the $n$-th subcarrier with $n \in \mathcal{N}_k$.
$\bm{p}=(\bm{p}_1,\ldots\bm{p}_K)$ denotes the vector of all transmit power vectors with the powers $\bm{p}_k = (p_{k1},\ldots p_{kn},\ldots p_{kN})$ for UE $k$. 
Lastly, $\underline{\bm{r}} = (\underline{r}_1,\ldots,\underline{r}_K)$ contains the target data rates of all UEs with the individual maximum total transmit power constraint $\overline{p}_k$ per UE $k$. 
\begin{align*} 
    \stackbin[\bm{p},\: (\mathcal{N}_k)_{k=1,\ldots K}]{}{\text{maximize}}  \quad &\mathlarger{\sum}_{k \in \mathcal{K}} \, \mathlarger{\sum}_{n \in \mathcal{N}_k} r_{kn} \\
    \text{subject to} \quad &\mathlarger{\sum}_{n \in \mathcal{N}_k} r_{kn} \, \geq \underline{r}_k \quad \forall \, k \in \mathcal{K} \\
    \quad &\mathlarger{\sum}_{n \in \mathcal{N}_k} p_{kn} \leq \overline{p}_k \quad \forall \, k \in \mathcal{K} \\
    \quad & \mathcal{N}_k \cap \mathcal{N}_j = \varnothing \quad \forall \, k,\,j \in \mathcal{K}, \quad k \neq j \\
    \quad & \bigcup_{k=1}^K \mathcal{N}_k = \mathcal{N} .
\end{align*}

Approaches like water filling, max-min fairness, weighted proportional fairness and utility optimization modify the given optimization problem in a manner to achieve either maximum throughput, maximum fairness or a tradeoff of both.
For further reading, we refer the interested reader to \cite{shamsSurveyResourceAllocation2014}.

\subsection{TimeFreqResourceAllocation-v0 Environment}
The TFRA environment is provided by Nokia's ``Wireless Suite'' problem collection \cite{NokiaWirelesssuite2021}.
It allows for a better comparability and reproducibility of research results by providing a set of standard environments against which to benchmark. 
The TFRA environment simulates an OFDMA downlink resource allocation task. 
An RL agent takes on the role of the scheduler that allocates a limited number of frequency resources, bundled into physical resource blocks (PRB), to a large number of UEs.
At each allocation step, the agent allocates one PRB to a UE. 
All available PRBs are allocated consecutively.
After all available PRBs have been allocated, one time step is completed.
The reward is composed of penalties for not satisfying the traffic requirements of the UEs and is hence always negative.
These requirements vary in GBR and PDB, depending on the QoS class of the UEs \cite{valcarceTimeFreqResourceAllocationv0Environment2020}.
The TFRA environment includes 4 different QoS classes, which are identified by their quality of service identifier (QI) $\mathfrak{q}_k \in \{1,2,3,4\}$.
As an environment is initialized, $K \in \{4\tilde{K}:\tilde{K}\in \mathbb{N}^+\}$ UEs are randomly spread over a \SI{1}{km^2} squared area. 
The area is an empty Euclidean space with a transceiver BS at its center.
The $K$ UEs roam around the square at constant speeds that are independently sampled from a normal distribution. 
The normal distribution parameters are chosen to emulate the speeds of pedestrians \cite{chandraSpeedDistributionCurves2013}. 
Bouncing off at the edges of the square at specular angles, the UEs move in random rectilinear trajectories. 
The scheduler has information about the age and size of all packets inside the BS buffer waiting for transmission.
Furthermore, the channel quality indicator (CQI) of every UE, which indicates the channel condition of a UE \cite{cqi_estimation}, as well as information about the QoS class of every UE is available at the scheduler.

\section{Reinforcement Learning}\label{chap:RL}
RL deals with the problem of learning how to skillfully map situations to actions in order to maximize a numerical reward signal.
The idealized form of the RL problem is an MDP.
Given uncertain and stochastic environments, MDPs are suitable to model most decision making problems \cite{luongApplicationsDeepReinforcement2019}.

A finite MDP is defined by a tuple $(\mathcal{S},\ \mathcal{A},\ p ,\ r)$, where $\mathcal{S}$ is a finite set of states and $\mathcal{A}$ a finite set of actions\cite{luongApplicationsDeepReinforcement2019}.
The dynamics of the MDP are fully described by the state-transition probability $p$ and the reward $r$. 
The state-transition probability $p(s_{t+1}|s_t,a_t): \mathcal{S}\times\mathcal{S}\times\mathcal{A}\rightarrow[0,1]$ gives the probability that an action $a_t \in \mathcal{A}$ taken on state $s_t \in \mathcal{S}$ results in follow-up state $s_{t+1} \in \mathcal{S}$.
The expected reward for a state transition described by a state-action triplet is a three argument function $r(s_t,a_t,s_{t+1}): \mathcal{S}\times\mathcal{A}\times\mathcal{S} \rightarrow \mathbb{R}$ resulting in a reward $r_{t+1} \in \mathbb{R}$ \cite[p.~48]{suttonReinforcementLearningIntroduction2018}.
Therefore, the state $s_t$, the action $a_t$ and the reward $r_{t+1}$ can be described by random variables $S_t$, $A_t$ and $R_{t+1}$.

In an MDP, the entity that makes the decisions and learns from interaction towards achieving a goal is called the agent. 
The agent continually interacts with an environment by selecting actions.
Depending on the state $s_t$ and the chosen action $a_t$ at time $t$, a reward $r_{t+1} \in \mathbb{R}$ is given to the agent.
To have an agent that not solely maximizes the instantaneous reward $r_{t+1}$, but chooses the actions $a_t \in \mathcal{A}$ to maximize future rewards, too, the discounted return 
\begin{align*}
    G_t \; \coloneqq \; R_{t+1} + \gamma R_{t+2} + \gamma^2 R_{t+3} + \cdots \; = \; \sum_{k=0}^{\infty} \gamma^k R_{t+k+1} \label{eq:dis_return} \, ,
\end{align*}
can be used, where $\gamma \in [0,1]$ \cite[p.~55]{suttonReinforcementLearningIntroduction2018}.

Any RL algorithm tries to find a policy $\pi \ : \ \mathcal{S}\rightarrow \mathcal{A}$ that maximizes the average return $\mathbb{E}\{G_t\}$.
A hypothetical optimal policy $\pi^*$ is defined as \cite{luongApplicationsDeepReinforcement2019}
\begin{align*}
    \pi^* = \arg \max_{\pi} \; \mathbb{E}\left\{\sum_{t=0}^{T} \gamma^t r\big( s_t, \pi (s_t ),s_{t+1} \big) \right\}.
\end{align*}
If an action $\pi(s_t)$ is taken on state $s_t$, the reward $r_{t+1}$ and the new state $s_{t+1}$ provide information to adjust the policy of the agent, repeating this process until the optimal policy is approached. 
One popular and effective method to obtain good policies in practice is the Q-learning algorithm \cite{luongApplicationsDeepReinforcement2019}.

The Q-function $Q(s,a)$ tries to approximate the expected discounted reward $G_t$ after taking an arbitrary action $a$ on an arbitrary state $s$ following policy $\pi$ \cite[p.~58]{suttonReinforcementLearningIntroduction2018}:
\begin{align*}
    Q(s,a) &\coloneqq \mathbb{E}^\pi\{G_t  \mid  S_t = s, A_t = a\} \quad \forall \, s \in \mathcal{S}, \, \forall \, a \in \mathcal{A}.   %&\,= \mathbb{E}^\pi \left[ \sum_{k=0}^{\infty} \gamma^k r_{t+k+1} \middle| s_t = s, a_t = a \right].
\end{align*}
Given state $s_t$, the expectation of $G_t$ for all possible actions $a_t \in \mathcal{A}$ can be estimated using the Q-function $Q(s=s_t,a=a_t)$. 
The action $a_t$ that maximizes $Q(s,a)$ is taken as the best possible action on state $s_t$.

To obtain an optimal policy, the Q-function for an optimal policy $Q^*(s,a):=Q_{\pi^*}(s,a)$ needs to be estimated, which is the objective of Q-learning \cite{watkinsQlearning1992}.
For all possible state-action pairs $(s_t,a_t) \in \mathcal{S} \times \mathcal{A}$, the optimal values of the Q-function need to be found \cite{luongApplicationsDeepReinforcement2019}. 
By observing the return $r_{t+1}$ of a state-action pair $(s_t,a_t)$, $Q^*(s,a)$ can be approximated using an iterative procedure by updating the Q-function as follows \cite[p.~131]{suttonReinforcementLearningIntroduction2018}:
%\begin{subequations}
    \begin{align*}
    Q(s_t, a_t) &\leftarrow Q(s_t, a_t)  \\
        &\quad + \alpha \Big[ r_{t+1} + \gamma \max_a Q(s_{t+1}, a) - Q(s_t, a_t) \Big] \nonumber \\
        &= (1-\alpha)Q(s_t,a_t) + \alpha \Big[ r_{t+1} + \gamma \max_a Q(s_{t+1}, a) \Big], \nonumber  
    \end{align*}
%\end{subequations}
where $\alpha \in [0,1]$ denotes the step size and $r_{t+1} + \gamma \max_a Q(s_{t+1}, a)$ is a more accurate state-action estimate that incorporates the observed reward information. 
This method adapts the Q-function in order to decrease the so-called temporal difference between the current value of the Q-function $Q(s_t, a_t)$ and the target value $r_{t+1}+\gamma \max_a Q(s_{t+1},a)$. 
Thus, an updated and more accurate Q-function on the given state-action pair is achieved.  
To ensure that previously unexplored states are explored during training and thus produce a larger reward in the long run, a random action is chosen with a probability of $\varepsilon \in [0,1]$ and $a \leftarrow \argmax_a Q(s_t,a)$ otherwise~\cite[pp.~26,100]{suttonReinforcementLearningIntroduction2018}. 

\begin{figure}
    \centering
    \includegraphics[]{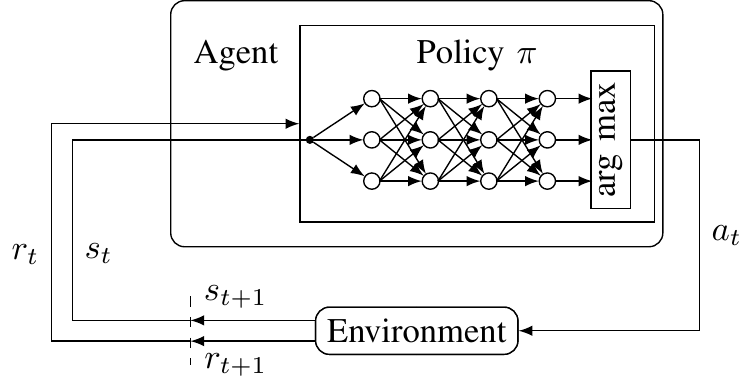}
    \vspace*{-1ex}
    \caption{The DQL agent interacting with the environment, adapted from \cite{luongApplicationsDeepReinforcement2019} and \cite[p.~48]{suttonReinforcementLearningIntroduction2018}. The DQN is utilized to derive the policy $\pi$.}
    \vspace*{-1ex}    
    \label{fig:mdp_dq}
  \end{figure}

For a small set of possible state-action pairs, a tabular Q-learning is feasible.
For a larger set it is limited by its inability to explore vast state-action spaces.
Deep Q-learning (DQL) overcomes this problem by using a DNN instead of a Q-table and learns an approximation of $Q^*(s,a)$, the deep Q-network (DQN) \cite{luongApplicationsDeepReinforcement2019}.
Figure \ref{fig:mdp_dq} shows the task of the DQN in an RL problem.

To update the parameters of the DQN in order to approach $Q^*(s,a)$, the backpropagation algorithm is applied.
For each state-action pair $(s_t,a_t)$ and its observed reward $r_{t+1}$, we use the Huber loss function between $r_{t+1}+ \gamma \max_a Q(s_{t+1}, a)$ and $Q(s_t,a_t)$. 
Instead of updating the DQN after every observed state-action pair, multiple tuples $(s_t,a_t,s_{t+1},r_{t+1})$ are observed and stored inside a \emph{replay memory}.
After a sufficient number of observations inside the replay memory, the backpropagation algorithm based on multiple samples of the replay memory, a \emph{mini-batch}, is applied, which yields smoother sample gradients.
To remove the correlation of successive state-action pairs inside the replay-memory and stabilize Q-learning, the state-action pairs used for backpropagation are randomly sampled from the replay memory \cite[p.440]{suttonReinforcementLearningIntroduction2018}.

\section{Deep Reinforcement Learning for OFDMA Downlink Resource Allocation}\label{chap:RAP_RL}
\subsection{Reward Design}
The success of learning a policy heavily depends on “how well the reward signal frames the goal of the application's designer” \cite[p.~469]{suttonReinforcementLearningIntroduction2018}.
For training our resource allocation agent, we use the reward $r^\text{(TFRA)}$ pre-implemented by the TFRA environment (see~\cite{valcarceTimeFreqResourceAllocationv0Environment2020} for details), however, we modify it to accelerate and kickstart training.
Inspired by kick-starting DRL training with help of expert agents \cite{schmittKickstartingDeepReinforcement2018} and expert learning \cite{wangDeepReinforcementLearning2019}, we introduce mimicking learning (MICKI).
We define the reward to be 
\begin{align*} 
    r_t = r_t^\text{(TFRA)} + r_t^{\text{(MICKI)}}
\end{align*}
with 
\begin{align*}
    r_t^{(\text{MICKI})} =
        \begin{cases}
         \mu(t), & \text{if} \ a_{t-1}^{\text{(DQN)}} = a_{t-1}^{\text{(expert)}}  \\
         0, & \text{otherwise,}
    \end{cases}
\end{align*}
where $\mu(t) \ : \ \mathbb{N} \rightarrow \mathbb{R}^+$ is a monotonously decreasing function with $\lim_{t \rightarrow \infty} \mu(t) = 0$.
We compare the action of our agent with the action choice of an expert agent running in parallel at the same state $s_t$.
If the same action is chosen, our agent achieves a bonus reward.
The value of the bonus reward decreases over time to encourage our agent to find ways to surpass the expert agent in performance. 
Owing to its simplicity, MICKI can be quickly implemented into any RL implementation where some expert agent is readily available. 
In our implementation, we choose a $\mu(t)$ that is constant within a training episode and decays exponentially with increasing training episodes. 

\subsection{User Equipment Shuffling}
During training of the agent, we observed that the agent often converges towards an agent which chooses the same action for every step.
This suggests that a UE bias creeps into the network, as the agent always chooses the same UE to allocate the PRB to.
As a result, the content of the replay memory is biased and therefore does not enable the agent to learn choosing other UEs than the one it is biased towards.
To remedy this problem, we randomly shuffle the order of all UEs in the data before feeding it to the DQN, reversing the shuffling at the network output
\begin{align*}
\underline{\bm{Q}} = \mathbf{P}_\text{rand}^{-1}\cdot \text {DQN}(\mathbf{P}_\text{rand} \bm{x}_\text{UEs})
\end{align*}
with $\bm{x}_\text{UEs} = \left(\bm{x}_1,\ldots \bm{x}_k, \ldots \bm{x}_{K}  \right)^T$, where $\bm{x}_k$ denotes the vector that contains all data of UE $k \in \mathcal{K}$.
$\mathbf{P}_\text{rand} \in \{0,1\}^{K \times K}$ denotes a random permutation matrix that shuffles the UEs.
$\mathbf{P}_\text{rand}$ is invertible which allows for a reversion of the shuffling at the output of the DQN.

\subsection{Encoder Neural Networks}
\begin{figure}
    \centering
    \vspace{-1cm}
    \includegraphics[]{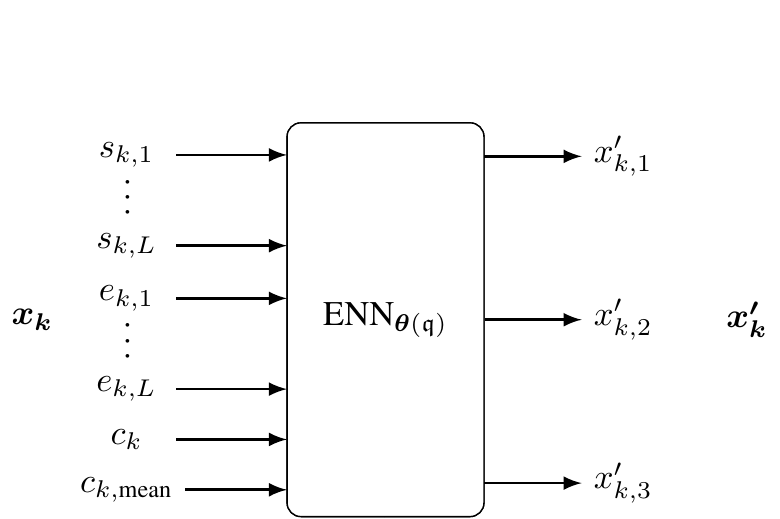}
    \caption{All UEs belonging to one QoS class use an encoder (ENN) with parameter set $\boldsymbol{\theta}(\mathfrak{q})$. The CQI of UE $k$ and its mean are denoted by $c_k$ and $c_{k, \text{mean}}$. $s_{k,l}$ and $e_{k,l}$ are the size and the age of packets in buffer slot $l \in\{1,2,...,L\}$ with a total buffer length of $L$. $\bm{x}'_k$ denotes the compressed information regarding UE $k$.}
    \label{fig:enc}
\end{figure}

To enhance the agent, full buffer state information can be used, see Sec.~\ref{chap:RAP}.
However, the dimensionality of the state vector quickly becomes prohibitively large and learning of the agent becomes difficult. 
According to \cite{murphyMachineLearningProbabilistic2012}, dimensionality reduction can often be employed to capture the “essence” of data and filter out inessential features.
Inspired by autoencoders \cite[p.~502]{goodfellowDeepLearning2016}, which are neural networks that are trained to copy their input to their output where a hidden layer describes a compressed representation of the input, we introduce ENNs.
All available state information of a UE is fed to an ENN to learn a compact representation of the state information.
The state information includes the CQI $c_k$ of UE $k$ and its mean $c_{k, \text{mean}}$ as well as the age $e_{k,l}$ and size $s_{k,l}$ of packets stored inside the BS's buffer at slot $l$, $l \in\{1,2,\ldots L\}$, where $L$ denotes the total buffer length.
To improve training of the ENNs for large sets of UEs, we instantiate four ENNs, where one ENN with parameters $\boldsymbol\theta(\mathfrak{q})$ is shared among all UEs with a QI $\mathfrak{q} \in \{0,1,2,3\}$.
Figure \ref{fig:enc} shows the setup of an ENN.
For a UE of the TFRA environment, we choose to reduce dimensionality from 66 DQN input features to 3 DQN input features. 

\subsection{Packet Shuffling}
We observed that during training the BS buffer rarely runs full and some buffer slots rarely ever contain a packet.
This imbalance of packet positions inside the BS buffer leads to an insufficient training of the ENN, since input neurons associated with rarely occupied buffer slots tend to have zero valued input during training.
To generalize for filled buffers, we introduce packet shuffling, where during training of the ENNs, the packet positions inside the BS buffer are shuffled.
For random packet shuffling (RPS) we randomly shuffle the packet positions so that packets are uniformly distributed along the buffer slots.
However, due to the random shuffling, the ENN cannot deduce the order of packets from their position anymore.
Therefore, we also introduce sorted packet shuffling (SPS), where we randomly shuffle the packet positions while preserving their respective order.

\subsection{Age capping}
Due to a limited training time for each training episode, the age $e_{k,l}$ of a packet assigned to UE $k$ and stored at buffer slot $l, \, l \in \{1,\ldots,L\}$, is limited by the duration of a training episode $T_\text{episode}$.
However, in deployment and validation, packets can have any age if they are not transmitted in due time.
As gradient descent optimizes the DQN parameters to minimize losses on the training data only, the network is unable to generalize to data that goes beyond the training set. 
Thus unpredictable behavior and limited robustness might occur in situations with large packet ages. 
We tackle this problem by capping packet ages by updating (``$\leftarrow$'') $e_{k,l}$ as
\begin{align*}
e_{k,l} \leftarrow \min \big(e_{k,l}\, , \; \text{PDB}(\mathfrak{q}_k) + 1 \big).
\end{align*}
We suggest that the packet age information is mainly of interest for estimating how soon the PDB of the packet will be exhausted.
Limiting packet ages to the PDB of the QoS class ensures that the DQN knows which packets have already exhausted their respective PDB.

\subsection{Embedding}
\begin{figure}
   \centering
   \scalebox{0.83}
   {\includegraphics[]{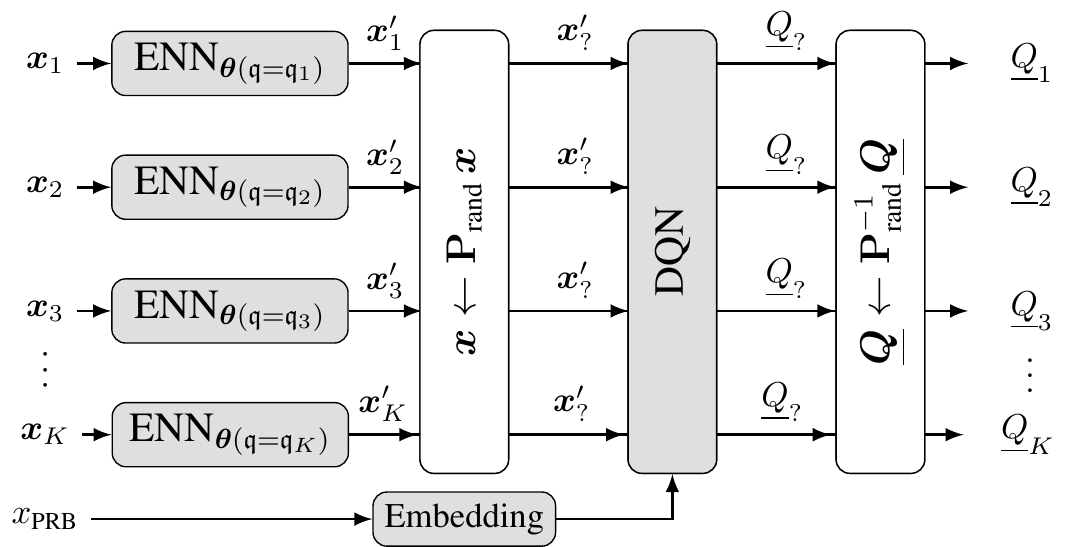}}
   \caption{The final structure of the agents network. All segments with learnable parameters $\boldsymbol\theta$ are highlighted in gray. The packet shuffling that is performed in some variants of the ENN is not pictured. For UE $k$ the ENN is chosen according to its QI $\mathfrak{q}_k \in \{1,2,3,4\}$.}
   \label{fig:full_net}
   \vspace*{-1ex}
\end{figure}

For each allocation step, categorical data $x_{\text{PRB}} \in \{1,\ldots,M\}$ that indicates the PRB to be assigned in the next step, is fed to the agent. 
To improve the meaningfulness of  $x_\text{PRB}$ for the DQN, it is translated to an $n$-dimensional vector by using learnable embeddings, which are parametrized lookup tables. 
We choose an embedding dimension $n = \left\lceil\sqrt[4]{N_\text{PRB}}\right\rceil$ \cite{tensorflowteamIntroducingTensorFlowFeature2017}.
For the TFRA environment with $N_\text{PRB} = 25$, $n=3$ and thus an embedding $\bm{x}_\text{PRB} \rightarrow \mathbb{R}^3$.
The final structure of the agent can be seen in Fig. \ref{fig:full_net}.

\section{Results}\label{chap:results}
\subsection{Experimental Setup}
For validation, the TFRA environment introduced in Sec. \ref{chap:RAP} is used.
$K=32$ UEs are initialized with 8 UEs per QoS class.
Before the TFRA environment proceeds by a physical time step, $N_\text{PRB}=25$ PRBs are allocated to the UEs.
The BS buffer for each UE contains $L=32$ buffer slots.
Furthermore, the set of TFRA environment initializations  is given by $\mathfrak{S}$.
To evaluate average training performance, we choose 7 environments to form the set $\mathfrak{S}_\text{training} \subset \mathfrak{S}$. 
We choose the duration of a training episode to be $T_\text{episode}=17\,500$ allocation steps, which corresponds to an environment simulation time of $\SI{700}{\milli \second}$.
After each 10 training episodes we validate the trained agents on an evaluation set $\mathfrak{S}_\text{eval} \subset \mathfrak{S}, \, \mathfrak{S}_\text{training} \cap \mathfrak{S}_\text{eval} = \varnothing$.
For final evaluation, we choose a test set $\mathfrak{S}_\text{test} \in \mathfrak{S}$ with $\mathfrak{S}_\text{test} \cap \mathfrak{S}_\text{training} = \varnothing$, $\mathfrak{S}_\text{test} \cap \mathfrak{S}_\text{eval} = \varnothing$ and $\vert \mathfrak{S}_\text{test}\vert =300$.
We evaluate four agents, which all use MICKI, UE shuffling and full BS buffer state information compressed by ENNs.
The ENN agent doesn't use additional techniques, while the no packet shuffling (NPS) agent additionally uses age capping.
The RPS and SPS agents use age capping and shuffle their packets randomly or sorted.
Table \ref{tab:tech_applied} shows the techniques applied to the agents.
We benchmark against the agents supplied by \cite{NokiaWirelesssuite2021}: the round robin if traffic (RRiT), proportional fair channel aware (PFCA) and the knapsack agent.
For benchmarking, we initialize $\mathfrak{S}_\text{test}$ and evaluate our agents against the benchmark agents for $T_\text{benchmark} = 65\,536$ allocation steps. 
Table \ref{tab:net_size_final} shows the dimension of the used NNs.
The obtained results are limited by a finite simulation time of the evaluation environments as well as a limited set of evaluation environments.

\begin{table}
    \centering
    \caption{The DQN network parameters, as used for the final trainings. }
    \begin{tabular}{l c c c}
        \toprule
        Parameter & Embedding & ENN & Main DQN\\
        \midrule
        input width & 25 (one-hot) & 66 & 99 \\
        output width & 3 & 3 & 32 \\
        depth & -- & 3 & 3 \\ 
        hidden widths & -- & (16,8) & (79,79)\\
        activation functions & linear & ReLU & ReLU, linear output \\
        \bottomrule
    \end{tabular}
    
    \label{tab:net_size_final}
\end{table}

\subsection{Learning performance}
\begin{figure}
    \centering
    \vspace{-0.5cm}
    \scalebox{0.19}
    {\includegraphics[]{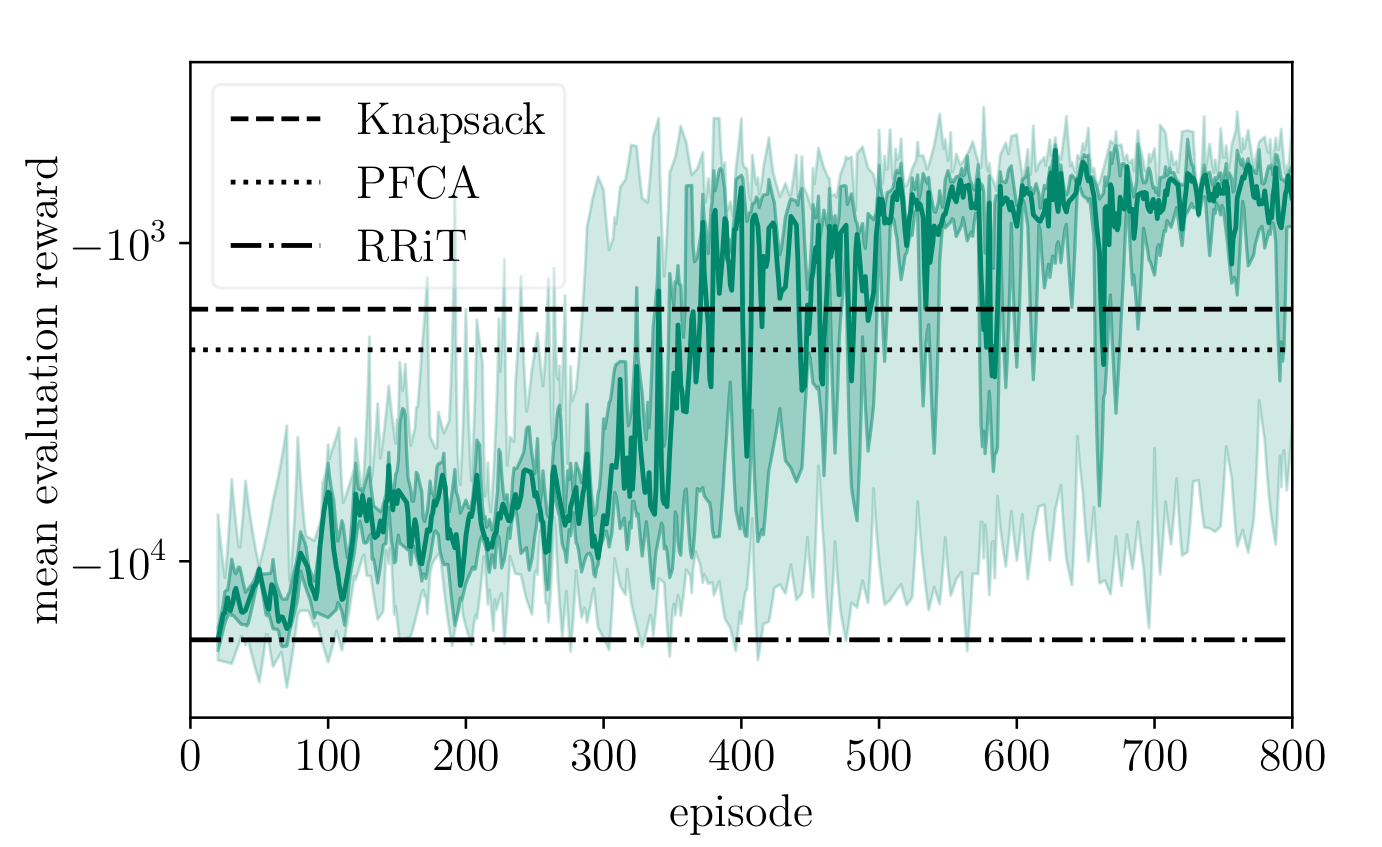}}
    \caption{Mean training evaluation rewards for the seven training runs of the RPS agent.}
    \vspace*{-2ex}
  \label{fig:fin_train_2}
\end{figure}

To evaluate training performance, we investigate the mean training evaluation rewards for seven training runs on $\mathfrak{S}_\text{eval}$, which can be seen in Fig. \ref{fig:fin_train_2} for the RPS agent.
Within an episode, the dark green line depicts the median evaluation value over all training runs. 
The second and third best values for each episode are averaged to obtain the upper limit of the dark shaded area, while the second and third worst values are averaged to obtain its lower limit. 
For each episode, the best and the worst evaluation value enclose the light shaded area. 
The dotted lines denote the performance of the knapsack, PFCA and RRiT agents when evaluated under identical settings.
We can show that during training, we outperform the benchmark agents on $\mathfrak{S}_\text{eval}$ and converge after a finite number of episodes.

\subsection{Performance evaluation}
\begin{figure}
  \centering
  \hspace{-5mm}
  \scalebox{0.65}{
  \includegraphics{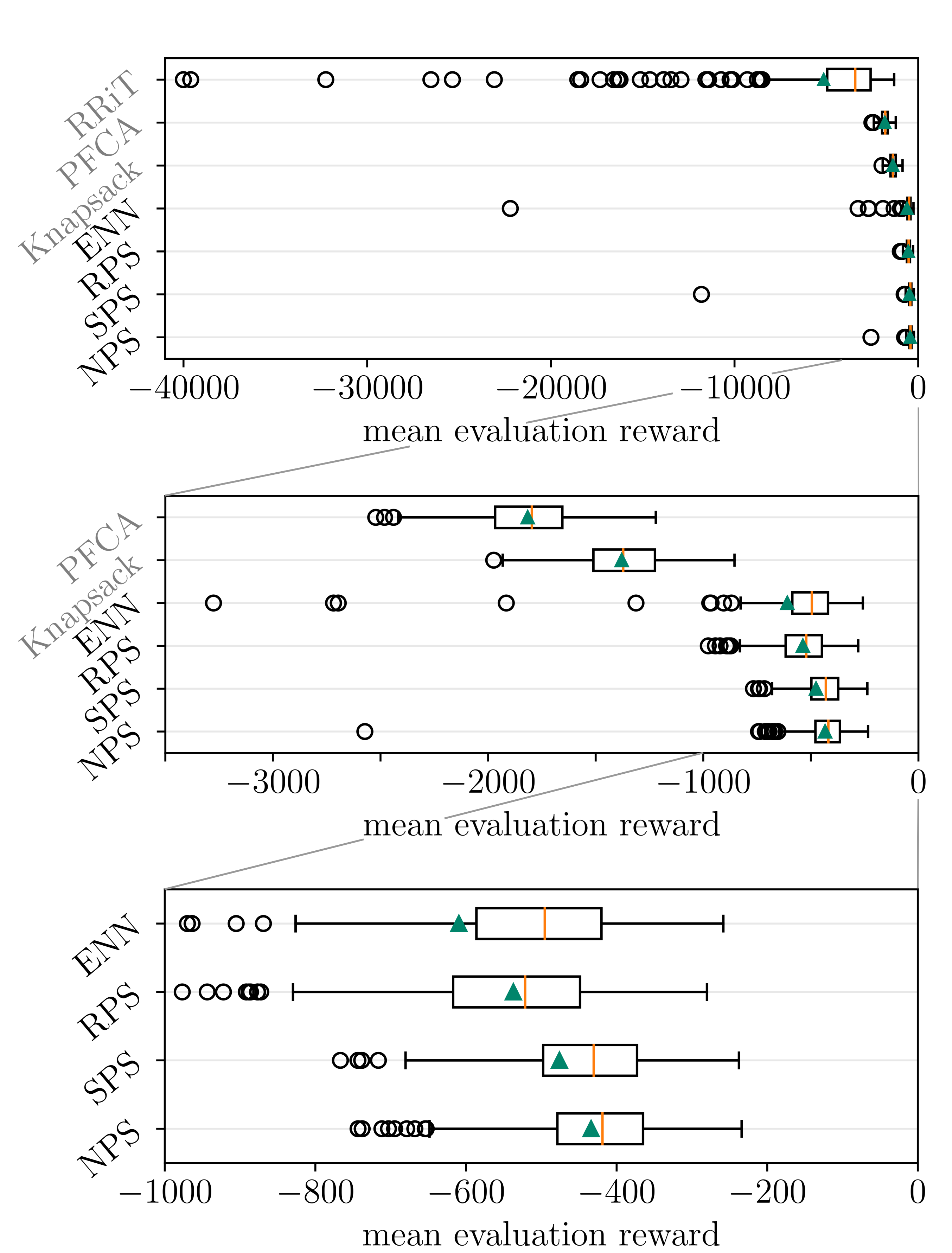}}
  \caption{Performance evaluation of our agents and the benchmark agents.}
    \label{fig:agent_result_boxplots}
\vspace*{-2ex}
\end{figure}

For each of the 300 environment initializations of $\mathfrak{S}_\text{test}$, we calculate the mean rewards achieved by the agents.
The distribution of evaluation rewards for the agents on different scales of the mean evaluation reward is shown in Fig. \ref{fig:agent_result_boxplots}.
The triangle indicates the mean of the evaluation reward and the vertical line inside the box, limited by the lower and upper quartile, depicts the median evaluation reward over all environments.

\begin{table}
  \centering
  \caption{Comparison of techniques applied to our agents.}
  \begin{tabular}{ccccc}
    \toprule 
     & ENN & NPS & RPS & SPS \\
    \midrule
    MICKI & $\surd$ & $\surd$ & $\surd$ & $\surd$ \\
    UE shuffling & $\surd$ & $\surd$ & $\surd$ & $\surd$ \\
    ENNs & $\surd$ & $\surd$ & $\surd$ & $\surd$ \\
    Age capping & --- & $\surd$ & $\surd$ & $\surd$ \\
    Packet shuffling & --- & --- & random & sorted \\
    \bottomrule
  \end{tabular}
  \label{tab:tech_applied}
\end{table}

All of our agents use full BS buffer state information and outperform the reference knapsack and PFCA agents in regards of median and mean performance by a factor of three.
The ENN agent has a large number of significant outliers while the NPS agent has only one significant outlier.
We believe that this improvement can be contributed to age capping and therefore the age capping technique generalizes an agent.
Compared to the SPS and NPS agents, we didn't observe any significant outlier for the RPS agent.
However, the SPS and NPS agents yield better mean and median rewards than the RPS agent.
We believe that using packing shuffling techniques we can trade generalization against performance.

\section{Conclusion}
In this work, we have proposed a setup and different learning techniques to train centralized RL agents for the OFDMA resource allocation problem with UEs that belong to different QoS classes.
We have proposed MICKI to improve upon the behavior of an expert agent.
To remedy the problem of learning a bias towards an action, we have introduced UE shuffling.
To outperform existing agents, we proposed ENNs to compress the most relevant information from the highly dimensional BS buffer states. 
Packet shuffling techniques can remedy issues caused by sparsely occupied BS buffers during training.
Since PDBs can only be exceeded to a limited degree during training, we have introduced age capping, which stabilizes the agent's performance for packet ages that far exceed the PDBs during deployment.
We have shown that our trained agents converge and outperform the benchmark agents supplied by the Nokia ``Wireless Suite''.
Our best agents outperform the benchmark agents by a factor of three.

\end{document}